\documentclass[aps,prd,twocolumn,showpacs,groupedaddress,nofootinbib,longbibliography]{revtex4-1}
\expandafter\let\csname equation*\endcsname\relax 
\expandafter\let\csname endequation*\endcsname\relax 

\usepackage{stmaryrd}

\usepackage{times}
\usepackage{amsfonts}
\usepackage{graphicx}
\usepackage{amsmath}
\usepackage{amssymb}
\usepackage{color}
\usepackage{subfigure}
\usepackage{natbib}
\usepackage{tensor}
\usepackage{accents}
\usepackage[colorlinks=true, citecolor=blue]{hyperref}
\usepackage{xspace}

 \DeclareMathOperator{\Tr}{Tr}

\def\beq{\begin{equation}}
\def\eeq{\end{equation}}
\def\bea{\begin{eqnarray}}
\def\eea{\end{eqnarray}}
\def\ben{\begin{enumerate}}
\def\een{\end{enumerate}}
\def\la{\langle}
\def\ra{\rangle}

\def\g{\gamma}
\def\d{\delta}

\def\z{\zeta}

\def\k{\kappa}
\def\l{\lambda}

\def\O{\Omega}

\def\r{\rho}

\def\L{{\cal L}}

\def\S{\Sigma }

\def\half{{\textstyle{\frac{1}{2}}}}

\def\cL{{\cal L}}

\begin{document}

\title{Entanglement Equilibrium and the Einstein Equation}

\author{Ted Jacobson}
\email{jacobson@umd.edu}
\affiliation{Kavli Institute for Theoretical Physics, University of California, Santa Barbara, CA 93106\\
Maryland Center for Fundamental Physics, 
University of Maryland, College Park, MD 20742}

\begin{abstract}

A link between the semiclassical Einstein equation and a maximal vacuum entanglement hypothesis is established. The hypothesis asserts that entanglement entropy in small geodesic balls is maximized at fixed volume in a locally maximally symmetric vacuum state of geometry and quantum fields. A qualitative argument suggests that the Einstein equation implies validity of the hypothesis. A more precise argument shows that, for first-order  variations of the local vacuum state of conformal quantum fields, the vacuum entanglement is stationary if and only if the Einstein equation holds. For nonconformal fields, the same conclusion follows modulo a conjecture about the variation of entanglement entropy.
\end{abstract}

\maketitle

\section{Introduction}

When restricted to one side of a spatial partition, the vacuum state of a quantum field has entropy because the two sides are entangled. The entanglement entropy of the restricted state is dominated by the ultraviolet (UV) field degrees of freedom near the interface, and hence scales with the area. This is similar to the Bekenstein-Hawking black hole entropy, $A/4L_p^2$, where $A$ is the horizon area and  $L_p = (\hbar G/c^3)^{1/2}$ is the Planck length  \cite{Bekenstein1972,Bekenstein1973,Hawking1974}. The similarity of these two ``area laws" is striking, and has led to the idea that black hole entropy is just a special case of vacuum entanglement entropy \cite{Sorkin1983,'tHooft1984,Bombellietal1986,FrolovNovikov1993,Srednicki1993,Solodukhin:2011gn}.
To match the Bekenstein-Hawking entropy, vacuum entanglement entropy should be cut off at the Planck scale.
Considering the gravitational backreaction of vacuum fluctuations, such a cutoff appears natural \cite{FrolovNovikov1993,Jacobson:2012yt}, but it lies deep in the regime of poorly understood quantum gravity effects.

Bekenstein defined the {\it generalized entropy} $S_{\rm gen}$ 
as the sum of the horizon entropy and the ordinary entropy in the exterior.
If the horizon entropy is indeed entanglement entropy, then the (fine-grained) generalized entropy is nothing but the total von Neumann entropy of the quantum state outside the horizon \cite{Sorkin:1986mg,Sorkin:1997ja,Solodukhin:2011gn}. Bekenstein
proposed the generalized second law (GSL) stating that 
$S_{\rm gen}$ never decreases \cite{Bekenstein1973}. The GSL has been shown to hold in various regimes \cite{Wall10proofs}, the proofs having been recently strengthened to apply to rapid changes and arbitrary horizon slices \cite{Wall:2010cj,Wall:2011hj}. The validity of the law depends on the Einstein equation, which relates the curvature --- and therefore the focusing of light rays that determines the change of horizon area --- to the local energy-momentum density of matter. 

The GSL thus points to a deep link between vacuum entanglement and the Einstein equation. 
The aim of this paper is to better understand the nature of this link. Motivated by the notion of 
vacuum as an equilibrium state, I formulate a maximal vacuum entanglement hypothesis (MVEH):
\begin{quote}\it
When the geometry and quantum fields are simultaneously varied from maximal symmetry, 
the entanglement entropy in a small geodesic ball is maximal at fixed volume. 
\end{quote}
%
This is formulated in the context of semiclassical gravity, i.e.\ quantum fields on a classical spacetime.
As such, it is predicated on the following assumption:
\begin{quote}\it
The area density of vacuum entanglement entropy $\eta$ is finite and universal.
\end{quote}
This assumption is supported by the evidence that horizon entropy can indeed be identified with entanglement entropy (see, e.g.\ \cite{Solodukhin:2011gn,Jacobson:2012ek,Cooperman:2013iqr}, and references therein). 
However, it involves UV aspects of quantum gravity that are not currently understood, so it remains an assumption.

I will argue that the Einstein equation supports the MVEH and, conversely, that the MVEH implies the
Einstein equation for first-order  variations of the local vacuum state for conformal fields.
For nonconformal fields the result holds modulo a conjecture about the variation of entanglement 
entropy to be explained below. 
It is well known that diffeomorphism invariance selects the Einstein equation, at second order in derivatives,
as the unique gravitational field equation in a metric theory.  
Since the MVEH is formulated in  a diffeomorphism-invariant fashion, 
it is therefore not surprising that the Einstein equation would arise. Nevertheless, entropy maximization
is quite different from Hamilton's principle of stationary action, so something
new is learned here. Moreover, the Newton constant that appears in the derived 
Einstein equation---which is {\it not} fixed by diffeomorphism invariance---has precisely the value required 
in order for $\eta$ to correspond to the Bekenstein-Hawking value, $1/4\hbar G$. This is a nontrivial and 
essential consistency property of the derivation.


Two lines of evidence motivated this paper.
First, the Einstein equation can be derived as a thermodynamic equation of state of the vacuum
outside a local causal horizon \cite{Jacobson:1995ab}. That derivation assumes that the entropy 
change of an otherwise stationary horizon is given by $\d Q/T$ when a local boost energy $\d Q$ 
crosses the horizon, $T=\hbar/2\pi$ being the Unruh temperature.
%
Second, recent work invokes AdS/CFT (anti-de Sitter/conformal field theory) duality, and the thermal nature of CFT vacuum entanglement entropy, to derive the linearized Einstein equation for perturbations of AdS spacetime \cite{Lashkari:2013koa,Faulkner:2013ica,Swingle:2014uza}. This approach treats the entropy statistically, rather than  
thermodynamically, and it concerns entropy of a compact region in the CFT at one time, rather than following the change of horizon entropy. The present work combines the local spacetime setting of the equation of state approach, with the 
statistical, compact region setting of the holographic analysis, but it proceeds directly in spacetime, 
making no use of holography.

\section{Area deficit and general relativity}

Einstein's field equation, 
\beq\label{Einstein}
G_{ab} = 8\pi G\, T_{ab},
\eeq
relates the Einstein curvature tensor $G_{ab}$ to the energy-momentum tensor of matter, $T_{ab}$.
Central to our story is the equivalence of \eqref{Einstein} to 
the statement that the surface area deficit of any small, spacelike geodesic ball of fixed volume 
is proportional to the energy density in the ball.\footnote{See Appendix \ref{AppA} for a related statement by Feynman.}
We begin by demonstrating this lovely relation. 

\begin{figure}
\centering
\includegraphics[scale=0.6]{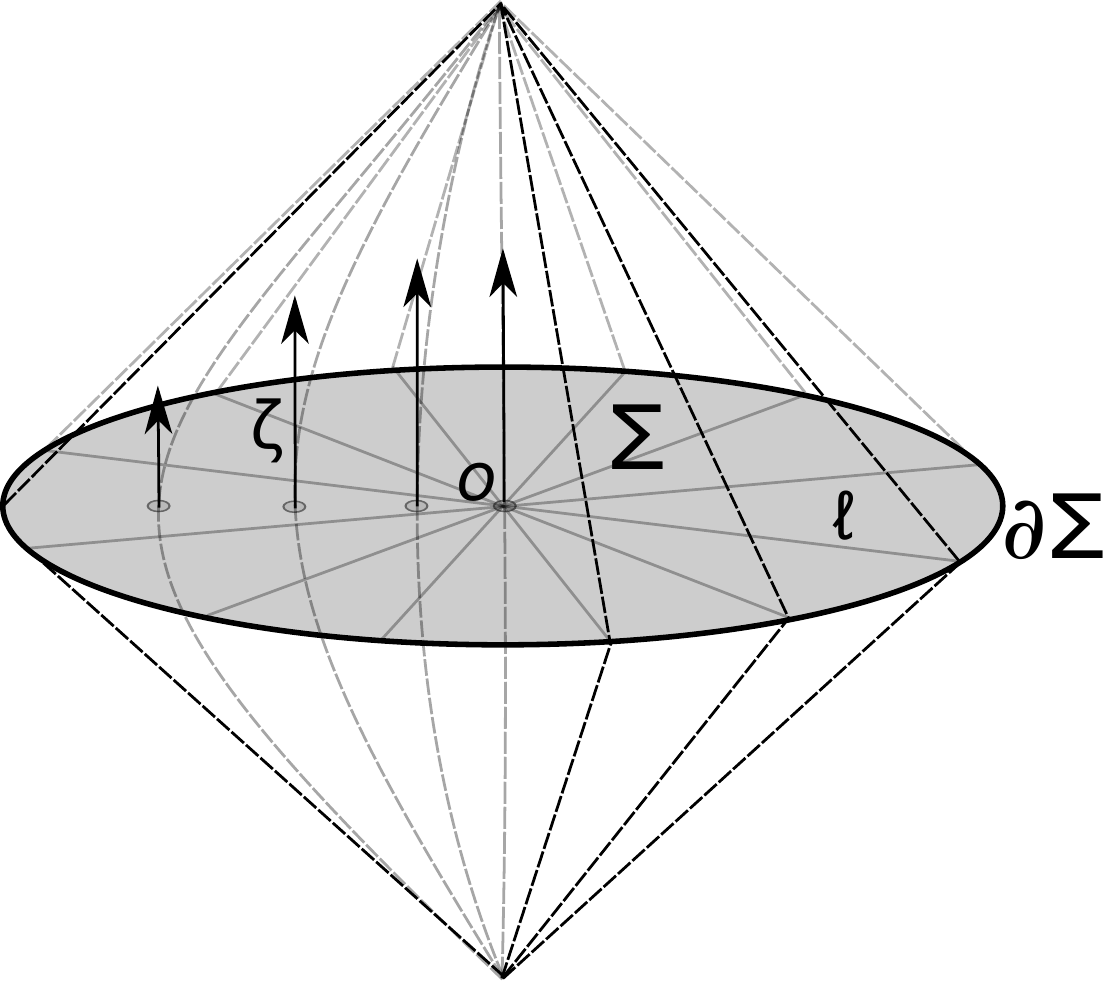}
\caption{Causal diamond, in a maximally symmetric spacetime, for a geodesic ball $\Sigma$ of radius $\ell$ with center $o$ and boundary $\partial \Sigma$. 
The dashed curves are flow lines of $\z$, the conformal Killing vector field, whose flow preserves the diamond
and  which vanishes at the top and bottom vertices and on $\partial\Sigma$.
The vectors show $\z$ at four points of $\Sigma$.}
\label{doublecone}
\end{figure}

At any point $o$ in a spacetime of dimension $d$, choose an arbitrary timelike unit vector $u^a$, and generate a $(d-1)$-dimensional spacelike ball $\Sigma$ by sending out geodesics of length $\ell$ from $o$ in all directions orthogonal to $u^a$. 
The point $o$ is the center of the ball, and the boundary $\partial \Sigma$ is the surface (see the grey region of Fig.~\ref{doublecone}). 
Choose a Riemann normal coordinate (RNC) system based at $o$, launched from an orthonormal basis formed by 
$u^a$ and $d-1$ spacelike vectors tangent to $\Sigma$.  Let the timelike coordinate be $x^0$, and
let the spacelike ones be $\{x^i\}$. The signature of the spacetime metric is taken here to be $({-}{+}{+}{+})$, and units are chosen with $c=1$.
%

We will assume the radius of the ball is much smaller than the local curvature length,
\beq
\ell\ll L_{\rm curvature}
\eeq
and work to lowest nontrivial order in their ratio. 
The volume variation at fixed radius, relative to flat space, is then given by
\beq\label{dV}
\d V|_{\ell} = -\frac{\O_{d-2}\ell^{d+1}}{6(d-1)(d+1)}{\cal R},
\eeq
where ${\cal R}=R_{ik}{}^{ik}$ is the spatial Ricci scalar  at $o$ (see Appendix \ref{AppA} for details),
and the area variation of $\partial\Sigma$ is given by $d\d V/d\ell$, i.e.\ 
%
\beq\label{dA}
\d A|_{\ell} = -\frac{\O_{d-2} \ell^d }{6(d-1)}{\cal R}.
\eeq
%
We will also be interested in the area variation at fixed volume, rather than at fixed geodesic radius.
When the radius of the ball varies, the volume and area variations have the additional contributions
$\d_r V = \ell^{d-2}\int \d r\, d\O$ and $\d_r A = (d-2)\ell^{d-3}\int \d r\, d\O$.
Choosing $\int \d r\, d\O$ so that the total volume variation vanishes, 
we obtain the area variation at fixed volume,
\beq\label{dAV}
\d A|_V = \d A - \frac{d-2}{\ell}\d V=
-\frac{\O_{d-2}\ell^d}{2(d^2-1)} {\cal R}. 
\eeq
This is smaller  by the factor 
$3/(d+1)$ than the variation at fixed radius \eqref{dA}.

To connect now with spacetime and the Einstein equation, note that the spatial Ricci scalar 
at $o$ is equal to twice the RNC $00$-component of the spacetime Einstein tensor: 
\beq\label{RG00}
{\cal R}=R_{ik}{}^{ik}=R-2R_0{}^0=2(R_{00}-\half R g_{00})=2G_{00}.
\eeq
The area deficit \eqref{dAV} can thus also be expressed as 
\beq\label{dAVG}
\d A|_V = -\frac{\O_{d-2}\ell^d}{d^2-1} G_{00}.
\eeq
%
Then, using the Einstein equation \eqref{Einstein}, 
we see that the area deficit is proportional to the energy density,
%
\beq\label{dAT}
\d A|_V = - \frac{8\pi G\O_{d-2}\ell^d}{d^2-1} \,T_{00}.
\eeq
%
Conversely, this
simple geometrical relation contains the full content of Einstein's equation,
if it holds at all spacetime points and for all timelike unit vectors. 

The evidence that the Einstein equation implies 
maximal vacuum entanglement can now be stated in a qualitative, intuitive fashion. 
Suppose the ball has a Bekenstein-Hawking entropy $A/4\hbar G$, arising from vacuum entanglement, and we try to increase
the entropy by placing an entangled qbit in the ball. 
To localize the qbit within a region of size $\ell$ we must give it an energy of
at least $\hbar/\ell$ which, according to \eqref{dAT}, will contribute an area deficit of order 
$\hbar G$, hence a surface entropy decrease of order unity, offsetting the
added qbit. It would not help to use a ``highly entropic object" with many internal states, because the existence of such objects makes its mark in
the vacuum as well, diluting the entropic effect of adding the object to the ball. Indeed, in the context of the Rindler wedge, it was argued that $\delta S \le \delta E/T$, where $T$ is the Unruh temperature $\hbar/2\pi$ and $\d E$ is the change of boost Killing energy, since a thermal state maximizes entropy at fixed energy \cite{Marolf:2003sq, Marolf:2004et}.
We now proceed to make this link between the Einstein equation and maximal vacuum entanglement more precise.

\section{Causal diamond and conformal isometry}

To evaluate the variation of the entanglement entropy in a spacelike geodesic ball $\Sigma$ 
it is helpful to consider the spacetime region causally determined by $\Sigma$,
called the {\it causal diamond} $D(\Sigma)$. 
In a maximally symmetric spacetime, $D(\Sigma)$
is the intersection of the future of a past vertex and the past of a future vertex,
and has a conformal isometry and rotational symmetry in the rest frame defined by these vertices
(see Fig.~\ref{doublecone}). 

The Minkowski line element $ds^2 = -dt^2 + dr^2 + r^2d\O^2$ takes the 
form $ds^2 = - du\, dv + r^2d\O^2$ with null coordinates $u=t-r$ and $v=t+r$. 
The Minkowski diamond centered on the origin consists of the intersection of the regions $u>-\ell$ and $v<\ell$. 
The unique conformal isometry that preserves the diamond, and is spherically symmetric, is 
generated by the conformal Killing vector 
%
\beq\label{zeta}
\zeta = \frac{1}{2\ell}\bigl[(\ell^2-u^2)\partial_u +(\ell^2-v^2)\partial_v\bigr]
\eeq
(for a derivation see Appendix \ref{AppB}).
Expressed in $t$ and $r$ coordinates, $\zeta$ is given by
\beq
\zeta = \frac{1}{2\ell}\bigl[(\ell^2-r^2-t^2)\partial_t -2rt\, \partial_r\bigr].
\eeq
The Lie derivative of the Minkowski metric along $\z$ is 
\beq\label{Lieg}
\L_\z \eta_{ab} = -(2t/\ell)\, \eta_{ab}.
\eeq
The vector $\z$ is tangent to the null generators on the past and future null boundaries of the diamond, 
so those boundaries are  conformal Killing horizons \cite{1979JMP....20..409D}. 
They meet at the ball boundary $\partial\Sigma$, where $\z$ vanishes,
so $\partial\Sigma$ is a bifurcation 
surface.  The surface gravity $\k$ of a conformal 
Killing horizon is well-defined by the equation $\nabla_a \z^2 = -2\k \z_a$ \cite{Jacobson:1993pf}, 
and with the normalization of $\z$ in \eqref{zeta} it is equal to unity. 

\section{Entanglement entropy of a diamond}


The entanglement entropy in a diamond $D(\Sigma)$ is the same as that in $\Sigma$.
Under a simultaneous variation of the geometry and the state of the quantum fields,
$(\d g_{ab}, \d|\psi\ra)$, the diamond entanglement entropy variation will consist of two 
contributions, a state-independent UV part 
$\d S_{\rm UV}$ from the area change induced by $\d g_{ab}$, and a state-dependent 
IR part $\d S_{\rm IR}$ from $\d|\psi\ra$. 

As mentioned above, we are assuming that, as a result of the UV physics, the entanglement entropy in a spatial
region is finite in any state, with a leading term $\eta A$. Here $A$ is the area of the boundary of the region and 
$\eta$ is a universal constant with dimensions [length]${}^{2-d}$.
The scaling with area is natural in any theory with a large density of states at short distances. The assumption that $\eta$ is universal is motivated by the idea that the UV structure of the vacuum is common to all states in the class being considered.\footnote{This involves an implicit choice of ``conformal frame" \cite{Flanagan:2004bz} for the metric, namely, the one for which $\eta$ is constant in spacetime. This metric turns out to satisfy the Einstein equation, so this frame is the so-called ``Einstein frame".}
Under this assumption, when the geometry is varied, the contribution to the entanglement entropy in 
$\S$ from the UV degrees of freedom near the boundary  $\partial\Sigma$
changes by an amount
\beq\label{SUV}
\d S_{\rm UV}=\eta\, \d A. 
\eeq
The total entropy variation will thus be given by 
\beq\label{dStot1}
\d S_{\rm tot} = \eta\, \d A + \d S_{\rm IR}
\eeq
If $\eta=1/4\hbar G$, \eqref{SUV} coincides with 
the variation of Bekenstein's generalized entropy, here interpreted as simply the total entropy in the diamond.
The MVEH implies that the total entropy variation \eqref{dStot1} is zero at first order , and negative for finite 
variations, when comparing to a maximally symmetric spacetime with 
the volume of $\S$ held fixed. 

To motivate this equilibrium condition,
we first recall that, 
for ordinary thermodynamic systems
in equilibrium, the Helmholtz free energy $F=E - TS$ is minimized at fixed volume.
The MVEH is analogous, but with the additional feature that the energy 
vanishes.
That the free energy of a diamond has no energy term can be motivated by 
comparison with de Sitter spacetime, and the restriction to fixed volume arises from 
the fact that the diamond has a conformal Killing vector rather than a true Killing vector (see Appendices \ref{AppC} and \ref{AppD}).

Our next step is to evaluate $\d S_{\rm IR}$.
The vacuum state of any QFT, restricted to the diamond, can be expressed (formally) as a
thermal density matrix, 
\beq\label{rK}
\r = Z^{-1}\exp(-K/T),\quad T=\hbar/2\pi,
\eeq
where $K$ is the ``modular Hamiltonian".  
The temperature $T=\hbar/2\pi$ is factored out here so that $K$ will be the 
generator of Lorentz boosts, i.e.\ hyperbolic angle shifts, at the edge of the diamond. 
For an infinite diamond that coincides with the Rindler wedge in Minkowski space, 
$T$ is the Unruh temperature \cite{Unruh1976, Sewell1982}.


Because $\r$ in \eqref{rK} has the form of a thermal state, it minimizes the modular free energy, 
\beq\label{FK}
F_K =\la K\ra - TS,
\eeq
where  the brackets denote quantum expectation value, and $S=-\Tr \r\ln\r$ is the von Neumann entropy. 
The variation $\d F_K$ must therefore vanish for any small variation  $\d\r$ of the state, i.e.\  
\beq\label{dSK}
\d S = \frac{2\pi}{\hbar}\d \la K\ra.
\eeq
This is just the usual Clausius relation for a ``thermal" state \eqref{rK}.

In general  $K$ in \eqref{rK} is not a local operator, and does not generate a geometric flow.
For a CFT, however, $K$ is equal to $H_\z$, the Hamiltonian generating the 
flow of the conformal boost Killing vector \eqref{zeta} \cite{Hislop:1981uh}. 
(This result is conformally related to the better-known version that holds for any
Poincar\'e invariant QFT restricted to the Rindler wedge \cite{Bisognano:1976za}.)
That is, $H_\z$ is given by the integral
\beq\label{Hz}
H_\z = \int_\Sigma T^{ab} \z_b\, d\Sigma_a.
\eeq
%
%
If the quantum field state is varied away from the vacuum, with an excitation length scale much 
longer than the diamond size, 
\beq\label{lexcitation}
\ell \ll L_{\rm excitation},
\eeq
then $\la T_{ab}\ra$ can be treated as constant, and 
using the Killing field \eqref{zeta}
we find
\beq\label{dHz}
\d \la H_\z\ra =\frac{\O_{d-2} \ell^d }{d^2-1}\, \d\la T_{00}\ra.
\eeq

If the matter field is {\it not} conformal, $K$ is not given by \eqref{Hz}, and we cannot directly use \eqref{dHz}.
However, suppose that the matter is described by a QFT with a UV fixed point, so it is 
asymptotically conformal at short distances, and that, in addition to \eqref{lexcitation}, 
the diamond is much smaller than 
any length scale in the QFT, 
\beq
\ell\ll L_{\rm QFT}.
\eeq
Then we conjecture---and we shall assume---that $\d \la K\ra$ 
has the form of  \eqref{dHz} with an additional term $\d X$ that is a spacetime scalar,
%
\beq\label{dKC2}
\d \la K\ra =\frac{\O_{d-2} \ell^d }{d^2-1}\, \bigl(\d\la T_{00}\ra+\d X\bigr).
\eeq
(The common coefficient is factored out to simplify later expressions.)
Calculations \cite{Casini:2016rwj,Speranza:2016jwt} indicate that for a class of theories and states,
this is the case, although in general 
$\d X$ may carry $\ell$ dependence and can dominate at small $\ell$.\footnote{In 
a previous draft of this paper, I had conjectured that $X=-\frac1d \la T\ra$, 
so that what appeared in $\d\la K\ra$ would be just the tracefree part of $\la T_{ab}\ra$.} 
Note that the relation \eqref{dKC2} refers only to the expectation value, and only to 
lowest order in the radius of the ball.

\section{Equilibrium and the Einstein equation}

We now postulate that a small diamond is in equilibrium if the quantum fields are in their vacuum state, and the curvature is that of a maximally symmetric spacetime (MSS) (Minkowski or (anti)-de Sitter). 
Any MSS seems an equally good candidate, so we will regard the curvature scale of the MSS as a local state parameter that is effectively constant in a small diamond but may 
depend on the diamond. 

The Einstein tensor in a MSS is $G^{\rm MSS}_{ab}=-\l g_{ab}$, with $\l$ a curvature scale. When the metric is varied away from the MSS, the area variation at fixed volume is obtained to lowest order in curvature by replacing $G_{00}$ in \eqref{dAV} with $G_{00}-G^{\rm MSS}_{00}$, which yields 
\beq\label{dAVGrel}
\d A|_{V,\l} = -\frac{\O_{d-2}\ell^d}{d^2-1} (G_{00}+\l g_{00}).
\eeq
The variation of the total diamond entropy \eqref{dStot1} 
away from the equilibrium can now be written using
\eqref{dAVGrel}, \eqref{dSK} , and \eqref{dKC2}:
\begin{align}
\label{dStot2}
\d S&_{\rm tot}|_{V,\l}=\eta\, \d A|_{V,\l}+\frac{2\pi}{\hbar}\d\la K\ra
= \frac{\O_{d-2}\ell^d}{d^2-1}\times\nonumber\\ 
&\left[-\eta \,(G_{00} +\l g_{00})+\frac{2\pi}{\hbar}\bigl(\d\la T_{00}\ra+\d X\bigr)\right].
\end{align}
The Einstein tensor should presumably be understood here as a quantum expectation value $\la G_{ab}\ra$, since the entropy that is maximized is, by definition, an expectation value.
In using \eqref{dKC2} for the matter entanglement variation, we are neglecting corrections that
would come from the curvature of the MSS, since those would be of higher order.

The requirement that the variation \eqref{dStot2} vanishes at all points and with all timelike unit vectors implies a tensor equation,
\beq\label{attempt1}
G_{ab} +\l\,g_{ab}= \frac{2\pi}{\hbar\eta}\bigl(\d\la T_{ab}\ra+\d X g_{ab}\bigr).
\eeq
The divergence of this equation, together with the Bianchi identity and local conservation of energy,  ties 
$\l$ to $\d X$ via
\beq\label{tie}
\l = \frac{2\pi}{\hbar\eta}\d X + \Lambda,
\eeq
where $\Lambda$ is a spacetime constant. 
Had we not allowed for the MSS curvature scale $\l$ in the equilibrium state,
\eqref{tie} would have implied the unphysical restriction that the scalar term $\d X$ be constant. Note also that
if $\d X$ has $\ell$-dependence then so does $\l$.

When \eqref{tie} is substituted back into \eqref{attempt1} we arrive at
\beq\label{bingo}
G_{ab} +\Lambda \,g_{ab}= \frac{2\pi}{\hbar\eta}\d\la T_{ab}\ra.
\eeq
This is Einstein's equation with an undetermined cosmological constant $\Lambda$, which evidently must 
be independent of $\ell$,
and with Newton's constant defined by 
\beq
G=\frac{1}{4\hbar\eta}.
\eeq
The area density of entanglement entropy $\eta$ and Planck's constant thus determine the gravitational coupling strength.
Stronger vacuum entanglement implies weaker gravity, i.e.\ greater spacetime rigidity. 
Note the crucial consistency: When expressed using $G$, the surface entropy $\eta A$ is the Bekenstein-Hawking entropy 
$A/4\hbar G$. The coefficient would have been off by the factor $(d+1)/3$ had we used the area variation at fixed
radius \eqref{dA} rather than at fixed volume \eqref{dAV}.

\section{Discussion}

We have shown, given our assumptions, that the semiclassical Einstein equation 
holds, 
for first-order  variations of the vacuum, 
if and only if the entropy in small causal diamonds is stationary at constant volume, when varied from a maximally symmetric vacuum state of geometry and quantum fields. 
We assumed the diamond size $\ell$ is much smaller than the local curvature length, 
the wavelength of any excitations of the vacuum, and
the scales in the matter field theory, but much larger than the UV scale at which quantum gravity effects become strong.
Our entanglement variation assumption for nonconformal matter \eqref{dKC2} 
concerns only standard QFT, and is either true or false. 

Strictly speaking the ``first-order  variation" refers to the derivative with respect to a parameter labeling the state, evaluated at the vacuum. To be physically applicable, however, the result should apply to finite but small variations. The example of a coherent state reveals a challenge in this regard \cite{Madhavan}: Such a state can have nonzero energy density while leaving entanglement entropy unchanged \cite{Fiola:1994ir, Benedict:1995yp}. That is, not all energy registers as a change of entanglement. This is consistent with the hypothesis of maximal vacuum entanglement, although the Einstein equation implies that the entropy has decreased --- relative to vacuum --- by more than it needs to in order to satisfy the hypothesis. Unless a further consequence of that hypothesis is found, or the hypothesis is refined and strengthened in some way, the Einstein equation does not appear to follow from it in all generality.

We close with some questions and remarks concerning the derivation and its implications.

\begin{itemize}

\item Do graviton fluctuations contribute to the entanglement entropy? 
The UV part of the entanglement entropy $S=\eta A$ is inscrutable  at this level,
and the IR part does not include gravitons. Since the diamond is taken much smaller than the wavelength of any ambient gravitons, 
they have no gauge-invariant meaning in the diamond. In the RNC gauge they are absent at first derivative order.
Moreover, the full, nonlinear Einstein tensor already appears on the geometric side of the equation, 
so it would be double counting to include any graviton energy. 

\item Can a gravitational field equation with higher curvature corrections be derived along these lines? Maybe.  
We neglected terms of order $\ell/L_{\rm curv}$ in the geometry calculations, whereas a next-higher-curvature correction to the field equation might be of order $(\ell_1/L_{\rm curv})^2$, where $\ell_1^2$ is the relative coefficient of the curvature squared term in the action. To capture this within our approximation would require $\ell/\ell_1<\ell_1/L_{\rm curv}$. The right-hand side is presumably less than unity, in order for higher-curvature terms to not dominate, so the diamond would have to be taken smaller than $\ell_1$.  If, say, $\ell_1$ were the string length,
would classical geometry and quantum field theory apply at that scale? Probably not. On the other hand, perhaps with improved accuracy of the geometric analysis, and the inclusion of subleading UV terms in the entanglement entropy, one could consistently capture higher-curvature corrections using a diamond larger than $\ell_1$.


\item A derivation of Einstein's equation invoking a quantum limit to measurements of the spacetime geometry of small causal diamonds was given in Ref.~\cite{Lloyd:2012du}. How are the assumptions used there related to those made here?


\item According to our derivation the Einstein equation is a property of vacuum equilibrium. 
Does this suggest how to include nonequilibrium effects? 

\end{itemize}

\vspace{3mm}
\acknowledgments

I am very grateful to H. Casini, W.~Donnelly, C.T.~ Eling, S.~Hollands, J.~Maldacena, D.~Marolf, M.~van Raamsdonk, V. Rosenhaus, C.~Rovelli, A. Satz, A.~Speranza, J. Suh, M. Varadarajan, and A.~Wall for helpful discussions, comments, and suggestions, and to anonymous referees for comments that led me to improve the presentation. This research was supported in part by the National Science Foundation under Grants No.\ PHY-1407744, and No.\ PHY11-25915.

\appendix

\section{Area deficit calculation}
\label{AppA}


For the Riemann normal coordinate (RNC) system based at a point $o$
we let $x^0$ denote the timelike coordinate and let $\{x^i=r n^i\}$ denote the 
spacelike coordinates, where $r$ is the geodesic distance and $n^i$ is a unit vector at $o$, 
$\d_{ij}n^i n^j=1$. The components of the metric at $o$ in this coordinate system are
$g_{00}=-1$, $g_{0i}=0$, and $g_{ij}=\d_{ij}$.  The spacelike geodesic ball of radius 
$\ell$ is denoted $\Sigma$. 
By definition, $\Sigma$ lies within the $x^0=0$ surface in this coordinate system. 
Using the standard result for the RNC metric components, 
the spatial metric $h_{ij}$ on $\Sigma$ takes the form 
\beq\label{metric}
h_{ij} = \d_{ij} -\tfrac{1}{3} r^2 R_{ikjl}\,n^k n^l + O(r^3)
\eeq
where $R_{ikjl}$ are the spatial components of the spacetime Riemann tensor evaluated at $o$.
The extrinsic curvature of $\Sigma$ vanishes at $o$, 
since $\Sigma$ is generated by geodesics from $o$. The components $R_{ikjl}$ are therefore 
also equal to the components of the spatial Riemann tensor. 

To lowest nontrivial order in the ratio $\ell/L_{\rm curvature}$
the volume element of $\Sigma$  is 
\beq\label{sqrth}
dV =\sqrt{h}\, d^{d-1}x =(1 - \tfrac16 r^2 R_{ik}{}^{i}{}_l\, n^k n^l)r^{d-2}dr\, d\O,
\eeq
where $d\O$ is the area element on the unit $(d-2)$-sphere.  
The integral over $d\O$ yields
\beq\label{intnn}
\int d\O \,n^kn^l = \frac{\O_{d-2}}{d-1}\d^{kl},
\eeq
where 
$\O_{d-2}$ is the area of the unit $(d-2)$-sphere. For spherically
symmetric integrands, the volume element is therefore 
\beq\label{sqrthr}
dV =\O_{d-2}\left(1 - \frac{r^2{\cal R}}{6(d-1)}\right)r^{d-2}dr,
\eeq
where ${\cal R}=R_{ik}{}^{ik}$ is the spatial Ricci scalar  at $o$.
Integrating \eqref{sqrthr} over $r$ from 0 to $\ell$ yields 
the volume variation at fixed radius written in the main text, 
\beq\label{dV}
\d V|_{\ell} = -\frac{\O_{d-2}\ell^{d+1}}{6(d-1)(d+1)}{\cal R},
\eeq
where ${\cal R}=R_{ik}{}^{ik}$ is the spatial Ricci scalar  at $o$.

In the main text, the relation \eqref{dV} is used to find the area variation
at fixed radius and at fixed volume, and the link to the Einstein
equation is made using the relation ${\cal R}=2G_{00}$, where 
$G_{ab}$ is the Einstein tensor.
The viewpoint here is not new. One can find 
the area and volume deficits at fixed radius, and the relation to the Einstein tensor,  
in Section 17 (``Riemannian coordinates and their applications") of Pauli's 1921 review of relativity \cite{PauliEncyc}.
 Pauli cites Riemann, as well as contemporary researchers, for the relevant geometrical relations.  He does not remark that the Einstein equation can be characterized purely in this manner. 
Feynman (in Section 11.2 of \cite{FeynmanGrav} and Section 42-3 of \cite{Feynman}) 
 expressed the Einstein equation for $d=4$ as the statement that, for all timelike directions,
the radius excess (equivalently the radius variation at fixed area) of a small sphere is given by  
$\ell - \sqrt{A/4\pi}=\d \ell|_A = GM/3c^2$, where $M=T_{00}V$ is the ``mass" contained in the sphere.
This can be derived from the relations above using 
[Eq. (4), main text]
together with 
$\d A|_\ell/A = -(d-2)\d \ell|_A/\ell$.

\section{Conformal isometry of a flat causal diamond}
\label{AppB}

The Minkowski line element $ds^2 = -dt^2 + dr^2 + r^2d\O^2$ takes the 
form $ds^2 = - du\, dv + r^2d\O^2$ with null coordinates $u=t-r$ and $v=t+r$. 
The diamond consists of the intersection of the regions $u>-\ell$ and $v<\ell$. 
To determine the conformal isometry that preserves the diamond, and is spherically symmetric, 
note that any vector field of the form 
\beq
\xi=A(u)\partial_u + B(v)\partial_v
\eeq
is a conformal isometry of the $du dv$ factor of the metric,  $\L_\xi du dv = [A'(u) + B'(v)]dudv$ (here $\L_\xi$ is the Lie derivative along $\xi$). It will be a conformal isometry of the full Minkowski metric provided
$\L_\xi r^2 = [A'(u) + B'(v)] r^2$.  Using $r=(v-u)/2$ we find that in fact $\L_\xi r^2 = (B-A)r$, 
so $\xi$ is a conformal Killing field if 
\beq
[A'(u) + B'(v)] (v-u)/2 = B(v)-A(u).
\eeq
At $u=v$ this implies $B(v)=A(v)$, hence at $v=0$ this condition becomes 
$[A'(u) + A'(0)] u/2 = A(u)-A(0)$. The general solution is 
\beq
A(u) = B(u) = a + bu + cu^2.
\eeq
The group generated by these vector fields is $SL(2,R)$.
To map the diamond onto itself, the flow of $\xi$ must leave invariant the boundaries $u=-\ell$ 
and $v=\ell$. This implies $A(\pm\ell)=0$, hence $A(u) = a(1-u^2/\ell^2)$.
Fixing the constant $a$ by requiring that $\xi$ have unit surface gravity, we thus obtain
the conformal Killing vector field 
\begin{align}\label{zetauv}
\zeta &= \frac{1}{2\ell}\bigl[(\ell^2-u^2)\partial_u +(\ell^2-v^2)\partial_v\bigr]\\
&= \frac{1}{2\ell}\bigl[(\ell^2-r^2-t^2)\partial_t -2rt\partial_r\bigr].\label{zetatr}
\end{align}
%

\section{Entropy of de Sitter space}
\label{AppC}

A small diamond 
is conformally isometric to a static patch of dS. 
The quantum statistical mechanics of dS was considered long ago by 
Gibbons and Hawking \cite{Gibbons:1976ue}. They examined the thermal partition function and argued that, since Euclidean dS is closed, the mass, angular momentum, and electric charge vanish, so that the only term in the free energy for pure dS space is the entropy term, $F_{\rm dS}=-TS$. The minimization of dS free energy is thus equivalent to the maximization of entropy. 

This result can also be understood via the ``first law of event horizons" \cite{Gibbons:1977mu}.
For a Schwarzschild black hole, this law would read $\d M = (\k/8\pi G)\d A + \d E_K$, where  $M$ is the mass, 
$\k$ and $A$ are the surface gravity and horizon area of the black hole, and $E_K$ is the Killing energy of a matter fluid 
surrounding the black hole \cite{Bardeen:1973gs}. 
For dS, instead the first law reads
\beq\label{dS1st}
(\k/8\pi G)\d A + \d E_K=0.
\eeq
There is no geometric total energy term since there is no asymptotic region. 
The vanishing of the 
total entanglement entropy variation $\eta\d A|_V + \d S_{\rm IR}$ for a causal diamond 
 is analogous to the first law of horizon mechanics 
in de Sitter space \eqref{dS1st}, identifying  $\d E_K$ with $\d S_{\rm IR}$, except for the 
restriction to fixed volume. This discrepancy can be traced to the fact that, unlike de Sitter space, 
a flat causal diamond has a conformal Killing vector rather than a true Killing vector.
In fact, one can derive a classical first law of flat causal diamonds which has the form \eqref{dS1st}, 
with $\d A$ replaced by $\d A|_V$, as we now demonstrate.

\section{First law of causal diamond mechanics}
\label{AppD}

Consider a conformally spherical causal diamond in a conformally flat solution to 
a diffeomorphism invariant gravity theory. 
There is a conformal Killing vector $\z$ that preserves the diamond, and for which the edge of the diamond is the bifurcation surface. We may invoke the variational identity 
\beq\label{1stlaw}
\d H_\z = \d \oint_{\partial\Sigma} dQ_\z,
\eeq
where $H_\z$ is the Hamiltonian generating evolution along the flow of $\z$ (constructed from the symplectic form evaluated on the Lie derivative of the fields along $\z$), $Q_\z$ is the associated
Noether charge $(d-2)$-form, and the variation is to any neighboring solution \cite{Iyer:1994ys}. 
The surface integral yields $-(\hbar\k/2\pi) \d S$, where $S$ is the Wald entropy associated with 
 the edge. 
(I include the factor $\hbar$ in the prefactor and the implicit $1/\hbar$ in the definition of the entropy, 
but these cancel in this strictly classical relation.)
If $\z$ were a Killing vector, $\d H_\z$ would vanish for all fields invariant under the Killing flow.
For matter treated as a fluid, it would not vanish, because the fluid potentials do not share the symmetry of the metric \cite{Iyer:1996ky,Green:2013ica}. Thus, for example, if the diamond were the static patch of de Sitter space in general relativity, we would recover in this way the first law of de Sitter horizons,
$(\k/8\pi G)\d A + \d E_\z=0$, where $\d E_\z$ is the matter Killing energy variation.

If $\z$ is only a conformal Killing vector,  $\d H_\z$ does not vanish. 
In general it receives contributions both from matter and from the gravitational field. 
If the theory is general relativity, and 
the background solution is Minkowski spacetime, then the gravitational field contribution to $\d H_\z$  
turns out to be $-(d-2)\k\d V/8\pi G\ell$ (see below).
When combined with the area variation term $-\k\d A/8\pi G$, this yields $-\k/8\pi G$ times 
the area variation $\d A|_V$ at fixed volume, [Eq. (5), main text].
Moreover, the conformal boost energy of the matter integrated over $\Sigma$ is given by the right hand side of 
[Eq. (8), main text],
when $T_{00}$ is constant. In this way we see that \eqref{1stlaw} recovers the statement of 
[Eq. (8), main text],
that addition of matter energy to the diamond decreases the area of the boundary at fixed volume.

To verify the result for $\d H_\z$ quoted above, 
we may use the expression for the symplectic form in \cite{Hollands:2012sf},
\beq\label{dHz}
\d H_\z = \frac{1}{16\pi G}\int_\Sigma w^a(g; \g_1,\g_2) d\Sigma_a,
\eeq
with
\beq
w^a = P^{abcdef}(\g_{2\,bc}\nabla_d\g_{1\,ef} - \g_{1\,bc}\nabla_d\g_{2\,ef}),\label{w}
\eeq
\begin{align}
P^{abcdef} &= g^{ae}g^{fb}g^{cd}-\half g^{ad}g^{be}g^{fc} -\half g^{ab}g^{cd}g^{ef}\nonumber\\
&~~~~~~~~~~~~~~~~~~~~~~-\half g^{bc}g^{ae}g^{fd}+\half g^{bc}g^{ad}g^{ef}.
\end{align}
In the case of interest the background metric $g_{ab}$ is the Minkowski metric $\eta_{ab}$,
$\gamma_2$ is the tangent vector $\cL_\z \eta_{ab}$ to the phase space flow generated by 
$H_\z$, and $\gamma_1$ is an arbitrary phase space tangent vector, i.e.\ a linearized solution about the background.
Now the conformal Killing vector in Minkowski coordinates is given by \eqref{zetatr}, so we have
\beq
\g_{2\,ef}=\cL_\z \eta_{ef} = -\frac{2t}{\ell} \eta_{ef}.
\eeq
Since this vanishes at $t=0$, only the $\nabla_d\g_{2\,ef}=-\frac{2}{\ell}\delta_d^t\eta_{ef}$ 
term of \eqref{w} contributes if the slice $\Sigma$ corresponds to the $t=0$ surface.
Then \eqref{w} yields
\beq
w^t = -\frac{d-2}{\ell}h^{bc}\g_{1\,bc} = -2\frac{d-2}{\ell}\frac{\d \sqrt{h}}{\sqrt{h}},
\eeq
where $h^{bc}=\eta^{bc}-u^b u^c$ is the metric on the spatial subspace orthogonal to $\partial_t$. 
Inserting this in \eqref{dHz} then yields
\beq\label{dHz2}
\d H_\z = -\frac{1}{8\pi G}\frac{d-2}{\ell}\d V.
\eeq
The surface gravity is unity for the chosen conformal Killing field, so this yields the result
claimed above. 

\bibliography{diamond}

\end{document}